\documentclass[prb,twocolumn,showpacs,floatfix,preprintnumbers,amsmath,amssymb,superscriptaddress]{revtex4}

\usepackage{graphicx}
\usepackage{dcolumn}
\usepackage{bm}
\usepackage{color}
\usepackage{color}
\usepackage[latin1]{inputenc}
\usepackage[urlcolor=blue]{hyperref}
\hypersetup{backref, colorlinks=true, linkcolor=blue, citecolor=blue}

\begin{document}

\title{Superconducting properties of sulfur-doped iron selenide}

\author{Mahmoud Abdel-Hafiez}
\affiliation{Center for High Pressure Science and Technology Advanced Research, Shanghai, 201203, China}

\author{Yuan-Yuan Zhang}
\affiliation{Key Laboratory of Polar Materials and Devices, East China Normal University, Shanghai 200241, China}

\author{Zi-Yu Cao}
\affiliation{Center for High Pressure Science and Technology Advanced Research, Shanghai, 201203, China}
\affiliation{Key Laboratory of Materials Physics, Institute of Solid State Physics, Chinese Academy of Sciences, Hefei 230031, China}

\author{Chun-Gang Duan}
\affiliation{Key Laboratory of Polar Materials and Devices, East China Normal University, Shanghai 200241, China}

\author{G. Karapetrov}
\affiliation{Department of Physics, Drexel University, Philadelphia, PA 19104, U.S.A.}

\author{V. M. Pudalov}
 \affiliation{P. N. Lebedev Physical Institute, Russian Academy of Sciences, Moscow 119991, Russia}
 \affiliation{Moscow Institute of Physics and Technology, Moscow 141700, Russia}

\author{V. A. Vlasenko}
 \affiliation{P. N. Lebedev Physical Institute, Russian Academy of Sciences, Moscow 119991, Russia}

\author{D. A. Chareev}
\affiliation{Institute of Experimental Mineralogy, Russian Academy of Sciences, 142432, Chernogolovka, Moscow District, Russia}

\author{O. S. Volkova}
\affiliation{Low Temperature Physics and Superconductivity Department, Physics Faculty, M.V. Lomonosov Moscow State University, 119991 Moscow, Russia}
\affiliation{Theoretical Physics and Applied Mathematics Department, Ural Federal University, 620002 Ekaterinburg, Russia}

\author{A. N. Vasiliev}
\affiliation{Low Temperature Physics and Superconductivity Department, Physics Faculty, M.V. Lomonosov Moscow State University, 119991 Moscow, Russia}

\affiliation{Theoretical Physics and Applied Mathematics Department, Ural Federal University, 620002 Ekaterinburg, Russia}
\affiliation{National University of Science and Technology "MISiS", Moscow 119049, Russia}

\author{Xiao-Jia Chen}
\email{xjchen@hpstar.ac.cn}
\affiliation{Center for High Pressure Science and Technology Advanced Research, Shanghai, 201203, China}

\date{\today}

\begin{abstract}
The recent discovery of high-temperature superconductivity in single-layer iron selenide has generated significant experimental interest for optimizing the superconducting properties of iron-based superconductors through the lattice modification. For simulating the similar effect by changing the chemical composition due to S doping, we investigate the superconducting properties of high-quality single crystals of FeSe$_{1-x}$S$_{x}$ ($x$=0, 0.04, 0.09, and 0.11) using magnetization, resistivity, the London penetration depth, and low temperature specific heat measurements. We show that the introduction of S to FeSe enhances the superconducting transition temperature $T_{c}$, anisotropy, upper critical field $H_{c2}$, and critical current density $J_{c}$. The upper critical field $H_{c2}(T)$ and its anisotropy are strongly  temperature dependent, indicating a multiband superconductivity in this system. Through the measurements and analysis of the London penetration depth $\lambda _{ab}(T)$ and specific heat, we show clear evidence for strong coupling two-gap $s$-wave superconductivity. The temperature-dependence of $\lambda _{ab}(T)$ calculated from the lower critical field and electronic specific heat can be well described by using a two-band model with $s$-wave-like gaps. We find that a $d$-wave and single-gap BCS theory under the weak-coupling approach can not describe our experiments. The change of specific heat induced by the magnetic field can be understood only in  terms of multiband superconductivity.
\end{abstract}


\maketitle
\section{Introduction}

The discovery of superconductivity with transition temperatures of up to 55\,K in iron-based pnictides has been at the forefront of interest over the last few years~\cite{Kam,XHC}. One of the most puzzling issues for these materials is the symmetry of the superconducting (SC) state~\cite{pag}. The pairing symmetry in Fe-based pnictides is under debate and various scenarios are being considered. Among these materials, iron selenide, FeSe, is of particular interest due to the following reasons: (i) it is considered as the simplest Fe-based superconductor~\cite{Thomale} for studying the pairing mechanism; (ii) in this system,  the static magnetism is missing and the structural ($\approx 90$\,K) and SC ($\approx 10$\,K) transition temperatures are well separated from each other~\cite{cwl}. From $^{77}$Se NMR measurements, the SC transition was found to  correlate with the enhancement of the spin fluctuations at low temperatures~\cite{Ima}; (iii) it is characterized by the absence of nesting between the hole and electron pockets of the Fermi surface~\cite{YMi}; (iv) the application of pressure around 9\,GPa leads to a strong enhancement of transition temperature ($T_{c}$) up to 37\,K~\cite{med}; (vi) in this system the largest SC  gap has been obtained by angle-resolved photoemission spectroscopy, which likely closes at 70\,K in extremely tensile strained FeSe~\cite{Peng}. The most mysterious property here is not even the pressure or strain induced $T_{c}$ increase (the cuprates have already shown the tendency of increased $T_{c}$ with reduction of the dimensionality), but a giant enhancement of the superconductivity at the Fe/SrTiO$_3$ interface, where SrTiO$_3$ (STO) has nothing in common with magnetic interaction. It seems that SrTiO$_{3}$ provides phonons that enhance superconductivity in single-layered FeSe~\cite{SJ}. Further transport measurements of the single FeSe/STO  has shown zero resistance state onset above 100\,K~\cite{JFG}, far above the liquid nitrogen boiling temperature.

Although FeSe system possesses many attractive features, the investigation of its physical properties is still in infancy. The material is composed of primarily PbO-type tetragonal FeSe$_{1-\delta}$ ($P$4/$nmm$) and partly of NiAs-type hexagonal FeSe ($P$63/$mmc$)~\cite{hsu}. The tetragonal structure is found to transform into an orthorhombic phase at low temperatures~\cite{jnm}. It remains unclear which of these phases is a superconducting one. It should be noted that the isotope effect experiments in Fe-based superconductors~\cite{IS}, show the iron isotope exponent ($\alpha _{c}$) values between 0.35 up to 0.4. Thus, one could infer that electron pairing in superconductors of the FeSe family is facilitated by  electron-phonon interaction. Furthermore, pure magnetic or spin-orbital interactions affect the interband coupling leading to decrease of thermodynamic $T_{c}$ like in the case of MgB$_{2}$. Identifying the origin of the SC pairing mechanism is the key to understanding these interesting properties of FeSe. There is no general consensus regarding the nature of pairing at the moment. For instance, the bulk probes, such as specific heat~\cite{JYLin}, Andreev reflections spectroscopy~\cite{DC}, thermal conductivity~\cite{JKD}, and the London penetration depth $\lambda _{ab}^{-2}(T)$~\cite{RKH,HA,RKH2} point to the existence of two-gap nodeless superconductivity. On the other hand,  evidence for nodal superconductivity in FeSe has been reported from the surface-sensitive probes, such as scanning tunneling spectroscopy~\cite{CLS}.

In this paper we report on magnetization, resistivity, London penetration depth, and low-temperature specific heat measurements of FeSe$_{1-x}$S$_{x}$. Although, similar investigations have been performed in detail on analogous compounds, $i.e.$ Fe(Se,Te)~\cite{HKi,RKH,HA,JHu}, such studies are lacking in the case of S-doped FeSe. Exploring the symmetry and structure of the order parameter, and the evolution of the SC gap with S doping in FeSe$_{1-x}$S$_{x}$ system through systematic measurements of temperature dependent specific heat and SC penetration depth is thus highly desired. In order to better understand the SC pairing mechanism it is  necessary to examine how these properties are affected by a different chemical composition. In the first part, we deal with magnetic measurements in magnetic \emph{dc} fields applied parallel to the $c$ axis. We show that the introduction of S to FeSe enhances the upper critical field $H_{c2}$, critical current density $J_{c}$, and the $T_{c}$. The upper critical fields $H_{c2}$ for $H$$\parallel$$c$ and $H$$\parallel$$ab$ have been determined from our detailed AC magnetization and specific heat studies, yielding a high superconducting anisotropy $\Gamma$ $\sim$ 4 for $x$ = 0.04. The anisotropy $\Gamma$ of the critical field is largest close to $T_c$ and decreases with decreasing temperature. From the measured temperature dependence of the specific heat, reliable values of the normal-state Sommerfeld coefficients are obtained for these materials. The second part of the paper is devoted to the study of the currently debated issue of the SC pairing symmetry by using high-quality single crystals of FeSe$_{1-x}$S$_{x}$. Based on the comprehensive low-$T$ measurements of the magnetic penetration depth and specific heat, we provide evidence for strongly-coupled multiband and nodeless superconductivity in FeSe family. In addition, the presence of multiple kinks in $\lambda^{-2}_{ab}(T)$ gives strong evidence for existence of two energy gaps in Fe(Se,S), which implies that several sheets of the Fermi surface contribute to the formation of Cooper pairs. Although the electronic specific heat of the SC state can be well described by using a two-band model, the change
of specific heat induced by a magnetic field can be understood only in terms of multiband superconductivity.

\begin{figure*}
\includegraphics[width=40pc,clip]{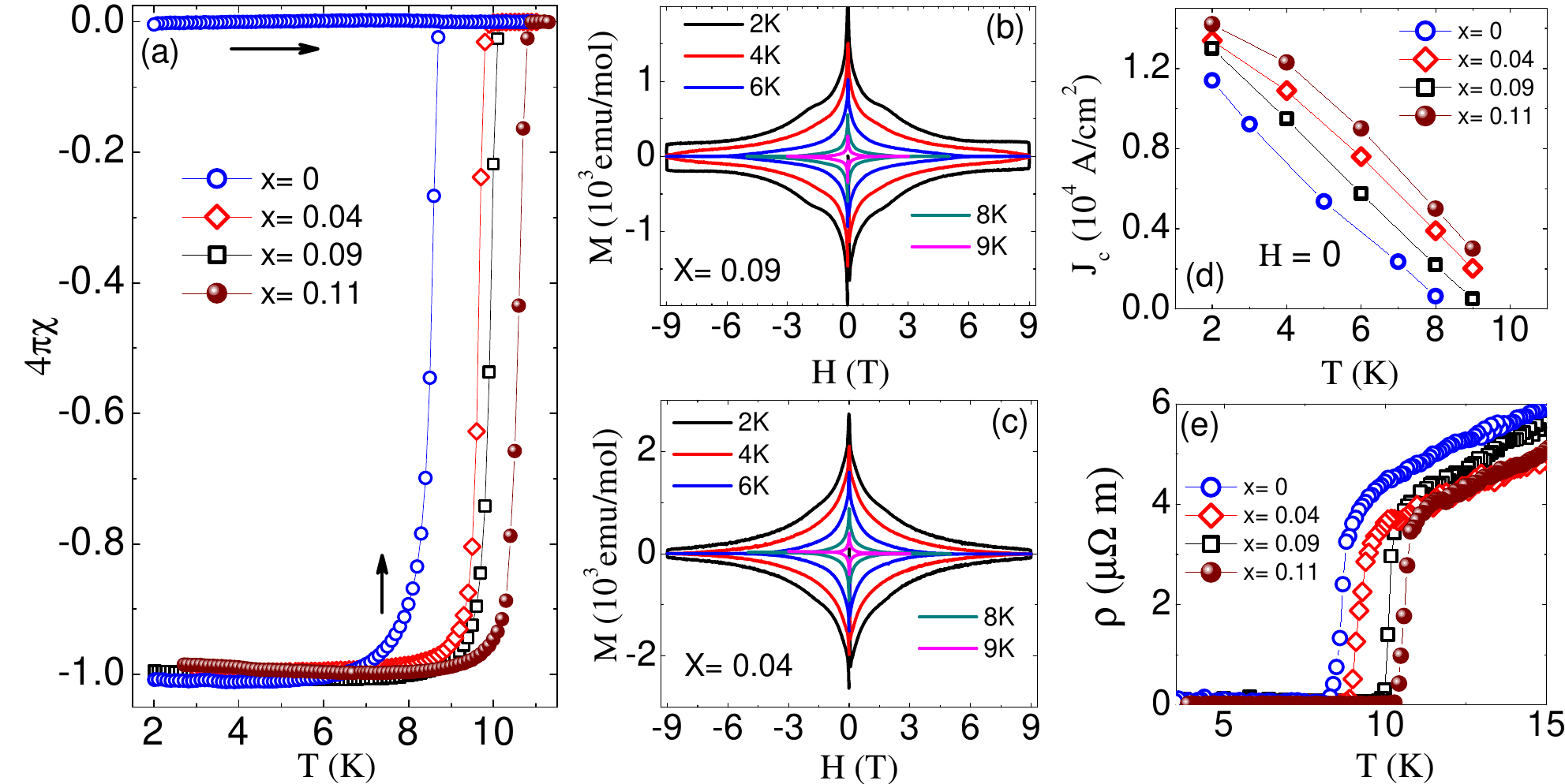}
\caption{\label{fig:wide} (a) presents the $T$-dependence of the magnetic susceptibility $\chi$ in an external field of 10\,Oe applied along the $c$-axis. $\chi$ has been deduced from the dc magnetization measured by following ZFC and FC protocols. (b) and (c) show the isothermal magnetization $M$ vs. $H$ loops measured at different temperatures ranging from 2 to 9\,K up to 9\,T applied along the $c$-axis. (d) illustrates the $T$-dependence of the critical current density $J_{c}$ values at $H$ = 0 for $x$ = 0, 0.04, 0.09 and 0.11. (e) shows the temperature dependence of the in-plane electrical resistivity in zero field up and represents a zoom of the resistivity data around the superconducting transition.}
\end{figure*}
\section{Experimental details}

All preparation steps like weighing, mixing, grinding and storage were carried out in an Ar-filled glove-box (O$_{2}$ and H$_{2}$O level less than 0.1 ppm). FeSe$_{1-x}$S$_{x}$ single crystals were grown in evacuated quartz ampoules using the AlCl$_{3}$/KCl flux technique in a temperature gradient (a hot part of the ampule at 400$^{o}$C and s cold part at 350$^{o}$C) for 45 days~\cite{DC}. The chemical composition of crystals was studied with a digital scanning electron microscope TESCAN Vega II XMU~\cite{DC}. The standard deviation of the average S concentration allows to judge upon the homogeneity of S within the crystals. Therefore, the composition and in particular the S-doping level was obtained by an average of over several different points of each single crystal. The analysis showed that the approximate chemical compositions are FeSe$_{1-\delta}$, Fe(Se$_{0.96\pm0.01}$S$_{0.04\pm0.01}$)$_{1-\delta}$, Fe(Se$_{0.91\pm0.01}$S$_{0.09\pm0.01}$)$_{1-\delta}$, and  Fe(Se$_{0.89\pm0.01}$S$_{0.11\pm0.01}$)$_{1-\delta}$. The crystals have a plate-like shape with the $c$-axis oriented perpendicular to the crystal plane. The crystals have only a tetragonal $\beta$-FeSe phase present. The lattice parameters $c$ = (5.52  $\pm$ 0.01)${\AA}$ and $a$ = (3.77 $\pm$ 0.01)${\AA}$ are found for FeSe single crystal. The quality of the grown single crystals was investigated by complementary techniques.

Magnetization measurements were performed by using a Quantum Design SQUID. The temperature dependent electronic transport was measured by using a standard four-probe alternating current dc method within a current applied parallel to the \emph{ab} plane. Electrical contacts parallel to the \emph{ab} plane were made using thin copper wires attached to the sample with silver epoxy. The low-$T$ specific heat was measured in the Quantum Design's Physical Property Measurement System within $T$ range from 2 to 14\,K in magnetic fields of up to $H$ = 9\,T applied along $c$ and $ab$-axis of the crystal. During the heat capacity measurements, the sample was cooled to the lowest temperature with an applied magnetic field [field cooled (FC)] and the specific heat data were collected using the adiabatic thermal relaxation technique.

\section{Results and discussions}

\subsection{Magnetization}

\subsubsection{DC magnetization measurements}

In Fig.\,1(a), we show the magnetic susceptibility $\chi$, measured with zero field cooling (ZFC) and field cooling (FC). $T_{c}$ has been determined from the onset of diamagnetic response to be around $\sim$ 8.5, 9.58, 10.1, and 10.7\,K for $x$ = 0, 0.04, 0.09, and 0.11 respectively. The FC and ZFC data show a sharp diamagnetic signal onset. The SC volume fraction of the crystals is close to 1, thus confirming bulk superconductivity and the high quality of the investigated systems. The clear irreversibility between FC and ZFC measurements is consequence of a strong vortex trapping mechanism, either by surface barriers or bulk pinning. The fact that the hysteresis loops are symmetric around $M=0$  line, points to relatively weak surface barriers and is indicative of strong bulk pinning~\cite{Pinn}. This consideration holds for all studied temperatures and investigated samples, even close to $T_c$ and indicates that vortex penetration occurs at a field close to the thermodynamic $H_{c1}$ (corrected by the demagnetization factor). Magnetization curves $[$Figs.\,1(b) and (c)$]$ show presence of a second peak for FeSe$_{0.91}$S$_{0.09}$ and FeSe$_{0.96}$S$_{0.04}$ for $H \parallel c$. The second peak effect has been studied extensively and its origin may be attributed to various mechanisms. The superconducting hysteresis loops $M(H)$ exhibits no paramagnetic background, which indicates that our investigated samples contain negligible amount of magnetic impurities and all Fe atoms are in nonmagnetic state of $Fe^{2+}$. From the irreversibility of the magnetization hysteresis loops in $M(H)$, we have extracted the magnetic field dependence of the critical current density $J_{c}$ at different temperatures (see Fig.\,1(d)). We used Bean's critical state model~\cite{Bean} in which the critical current is constant across the sample and the critical current density in a platelet sample is given by:

\begin{figure*}
\includegraphics[width=40pc,clip]{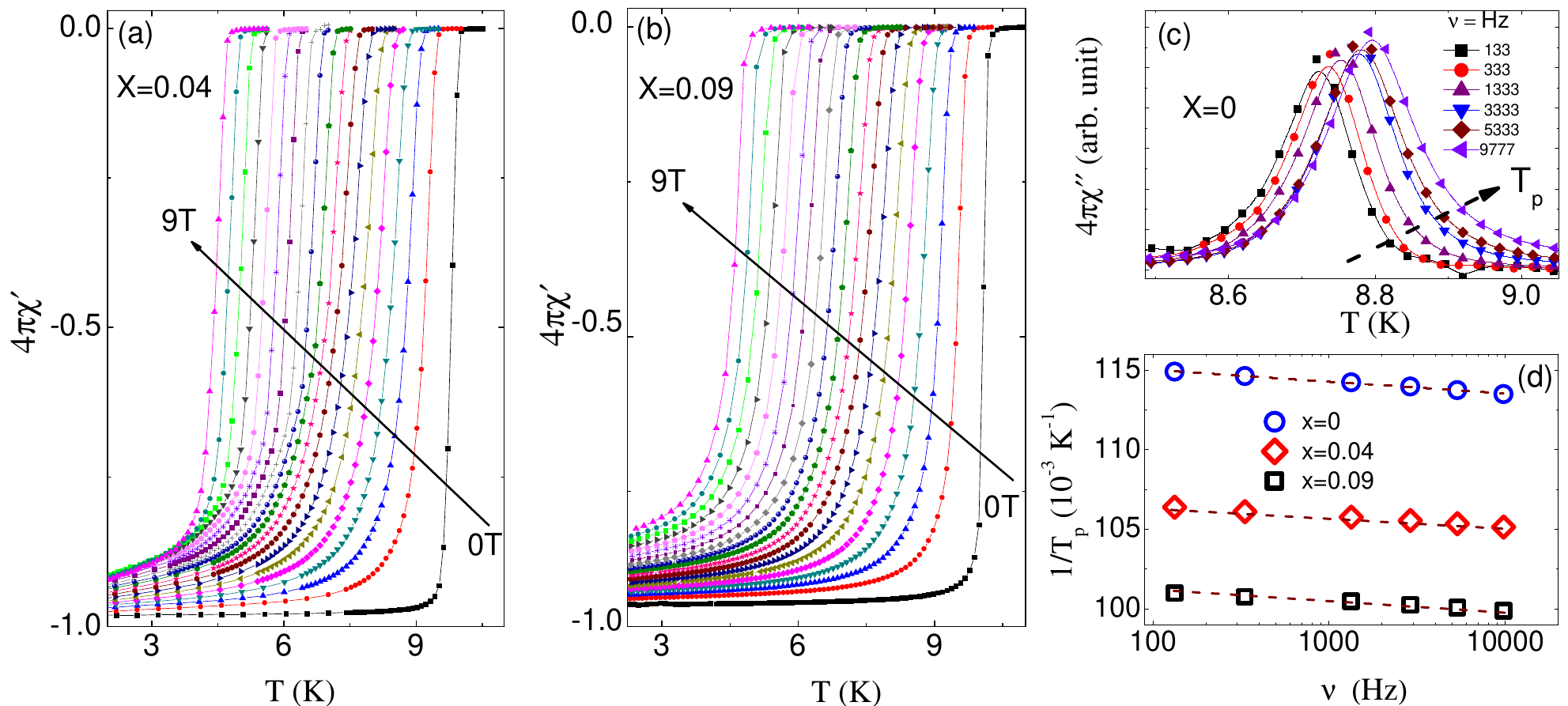}
\caption{\label{fig:wide} (a) and (b) summarize the temperature dependence of the complex AC-susceptibility components $4\pi\chi{'}_{v}$ of $x$ = 0.09, and 0.11 measured in an AC field with an amplitude of 5\,Oe and a frequency of 1\,kHz. The data were collected upon warming in different DC magnetic fields after cooling in a zero magnetic field. (c) presents the imaginary part of AC at various $\nu _{m}$ for $x$ = 0. (d) shows the values from the position of the maxima of the imaginary part in (c) vs. $\nu _{m}$ for $x$ = 0, 0.04 and 0.09.}
\end{figure*}

\begin{equation}
\label{eq2}
J_{c} = \frac{20\Delta M}{[a(1-\frac{a}{3b})]},
\end{equation}
where $\Delta M$ = $M_{dn}-M_{up}$, $M_{dn}$ and $M_{up}$ are the magnetization values measured on the decreasing and increasing branches of M(H), respectively, $a$ [cm] and $b$ [cm] are sample sizes perpendicular to the applied field ($a < b$). Here $\Delta M$ is in electromagnetic units per cubic centimeter and the calculated $J_{c}$ is in Ampere per square centimeter. The calculated $J_{c}$ values are summarized in Table I. Figure 1(e) shows the in-plane resistivity data for $x$ = 0, 0.04, 0.09, and 0.11 samples near $T_{c}$. A sharp SC transition is seen in all of the samples which is in agreement with the magnetization data  in Fig.\,1(a). Upon cooling the resistivity decreases monotonically and shows a metallic behavior.

\subsubsection{AC magnetization measurements}

Figures 2(a) and (b) depict the temperature dependence of the volume AC susceptibilities $\chi{'}_{v}$ for $x$ = 0.04, and 0.09 respectively. The measurements were done in an AC field with an amplitude $H_{AC}$ = 5\,Oe and a frequency $f$ = 1 kHz in DC fields up to 9\,T parallel to the $c$ axis. Special care has been taken to correct the magnetization data for demagnetization factor, where the demagnetization factor has been estimated based on crystal dimensions \cite{Osborne1945}. In general, AC-susceptibility measurements can be used for an investigation of the flux dynamics in superconductors. The imaginary part $\chi{''}_{v}$ is related to the energy dissipation in the sample due to vortex motion and the real part $\chi{'}_{v}$ is related to the amount of Meissner currents screening. Both functions depend on the ratio between the skin depth $\delta_s$ and the sample dimension $L$ in the direction of the flux penetration. In the normal state $\delta_s\sim(\rho_n/f)^{0.5}$, where $\rho_n$ is the normal-state resistivity and \emph{f} is the frequency \cite{M1}. In the superconducting state, the skin depth $\delta_s\propto\lambda_L$ if an external magnetic field is below the first critical field $H_{c1}$, where $\lambda_L$ is the London penetration depth. For magnetic fields above $H_{c1}$, $\delta_s\propto L_B$, where $L_B\sim\ B_{ac}/J_c$ is the Bean's penetration depth and $J_{c}$ is the critical-current density. In general, if $L\ll\delta_s$ the AC field penetrates completely the sample, although the susceptibility is small. In the opposite case, $L\gg\delta_s$, most of the sample volume is screened. Therefore, $4\pi\chi{'}_{v}=-1$ and $\chi$$''$$_{v}$ $\rightarrow$ 0.

\begin{figure*}
\includegraphics[width=40pc,clip]{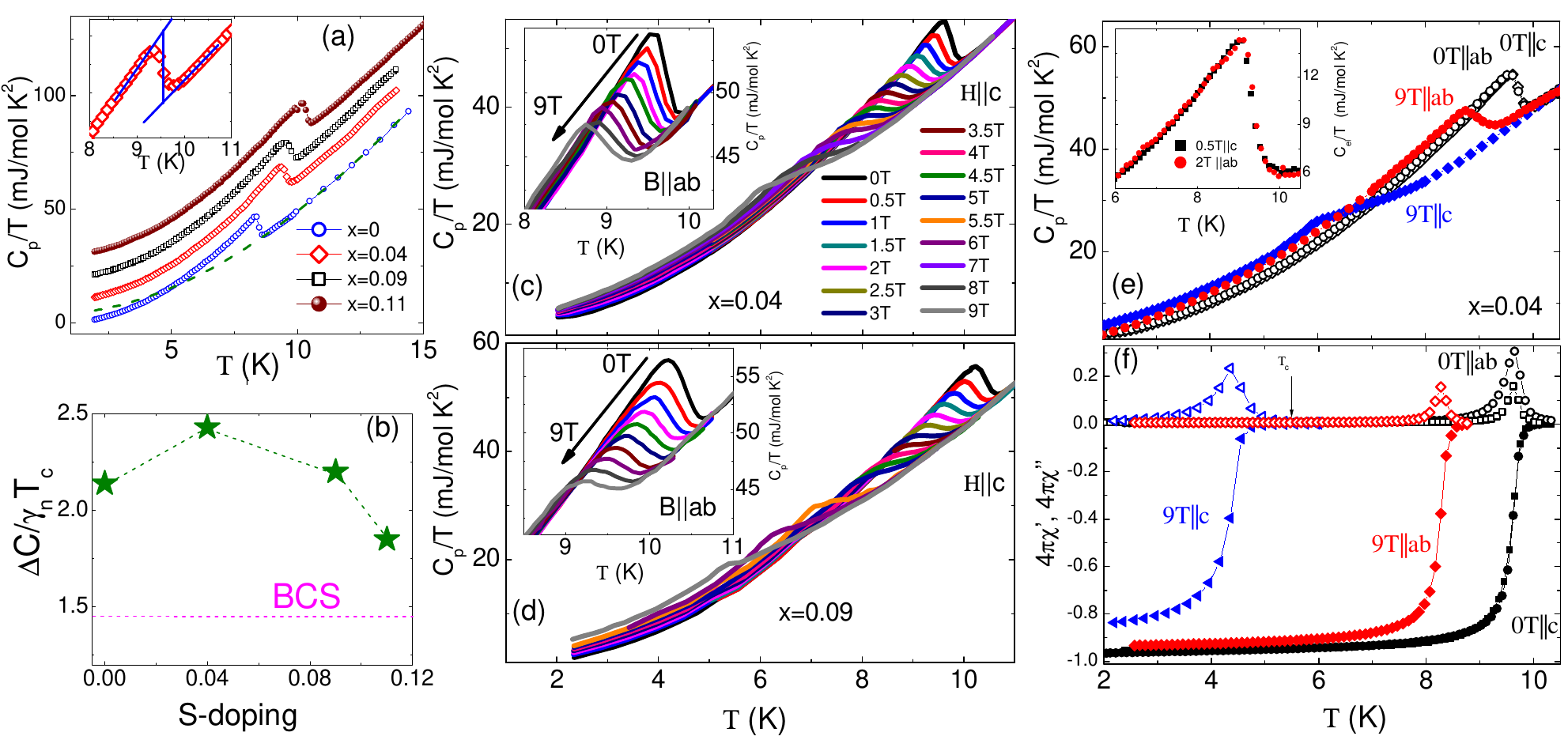}
\caption{\label{fig:wide} (a) $C_{p}/T$ vs. $T$ of $x$ = 0, 0.04, 0.09, and 0.11 in zero magnetic field. Data are shifted by an offset along the $y$ axis for clarity. The upper inset: enlarged  $C_{p}/T$ vs. $T$ plot near the SC transition for $x$ = 0.04. The lines illustrate how $C_{p}/T_{c}$ and $T_{c}$ are estimated. The dashed line is the fitting of the specific heat below 14\,K by using $C_\mathrm{p} = \gamma _{n}T + \alpha T^3 + \beta T^5$. (b) The doping dependence of the normalized specific heat jump is $\Delta C_{p}/\gamma _{n}T_{c}$ for $x$ = 0, 0.04, 0.09, and 0.11. (c) and (d) summarize the temperature dependence of the specific heat of $x$ = 0.04, and 0.09 respectively in various applied magnetic fields up to 9\,T parallel to the $c$ axis and parallel to the $ab$ plane as presented in the insets. (e) presents the temperature dependence of the specific heat of both orientation for 0\,T and 9\,T two data sets with the same $T_{c}$ value for the two directions. The inset highlights the electronic specific heat data after subtracting the phonon contribution for $x$ = 0.04 of two data sets with the same $T_{c}$ value for the two directions. The data present the 0.5\,T $B \parallel c$ and for 2\,T $B \parallel ab$. (f) the temperature dependence of the complex AC-susceptibility components $4\pi\chi{'}_{v}$ and $4\pi\chi{''}_{v}$ of $x$ = 0.04 of both orientation for 0\,T and 9\,T measured in an AC field with an amplitude of 5\,Oe and a frequency of 1\,kHz. The data in (e) and (f) show that the electronic specific heat divided by temperature for $B \parallel ab$ and $B \parallel c$ and the AC-susceptibility represent an anisotropy of $\Gamma$ = 4 for S-doping concentration.}
\end{figure*}

In accordance with this, the AC-susceptibility data measured at low temperatures confirm the bulk superconductivity of the investigated crystals. The transition temperature $T_{c}$ has been extracted from the bifurcation point between $\chi{'}_{v}$ and $\chi{''}_{v}$. This point is related to the change in the resistivity due to the superconducting transition. It can be also used for the determination of the temperature dependence of the upper critical field $H_{c2}$ from the AC-susceptibility data measured at various DC fields. Therefore, the most natural way to investigate the vortex dynamics is to repeat $\chi _{ac}$ vs. $T$ scans at a fixed $H_{dc}$ at different frequencies $\nu _{m}$ in order to employ empirical peak functions around the maxima. Figure 2(c) presents the temperature dependence of the imaginary part of AC susceptibility at various frequencies $\nu _{m}$ for FeSe. One can clearly see that the peak maxima shifts to a higher temperatures upon increasing the frequency which is apparently due to the motion of vortices. Figure 1(d) shows $T_{P}$ values, the position of the maxima of imaginary part in (c) vs. $\nu _{m}$. One can notice that, similarly to what was observed in YBa$_{2}$Cu$_{3}$O$_{7}$~\cite{G1} and CeFeAsO$_{0.92}$F$_{0.08}$~\cite{G2}, the quantity 1/$T_{p}$ presents a frequency  dependence over the explored range of $\nu$ the explored frequency range at fixed applied field $H$ $[$$H_{ac}$ = 1\,Oe and $H_{dc}$ = 1\,T, see Fig.\,2(d)$]$. The latter phenomenology is well verified for all the samples $x$ = 0, 0.04 and 0.09. The dashed line outlines the expected logarithmic trend typical for thermally-activated process according to: $1/T_{p}(\nu _{m}) = - \frac{1}{U_{0}}\ln\frac{\nu _{m}}{\nu _{0}}$. The parameter $\nu _{0}$ represents an intraband condensate vortex hopping. From the latter equation, it can be observed that the logarithmic behavior of 1/$T_{p}$ is mainly controlled by the parameter $U_{0}$, playing the role of an effective depinning energy barrier in a thermally activated flux creep model.

\subsection{Specific heat}

Low-temperature specific heat $C_{P}$, being equal to the temperature derivative of the entropy $S$, and probes the gap structure of bulk superconductors. The thermodynamic $C_P$ measurements well complement the magnetic ($\lambda$) measurements, since the former is hardly affected by vortex pinning. The information about the pairing symmetry is contained in the $C_{el}$, which is proportional to the quasiparticle density of states (DOS) at the Fermi energy. First we address the zero-field $T$-dependent
specific heat data of FeSe$_{1-x}$S$_{x}$ plotted as $C_{p}/T$ vs $T$ (Fig.\,3(a)). A clear sharp anomaly is observed due to the SC phase transition. In order to determine the specific heat related to the SC phase transition we need to estimate the phonon ($C_\mathrm{ph}$) and electron ($C_\mathrm{el}$) contributions. At low temperatures, $C_\mathrm{el}$ behaves linearly with temperatures, while $C_\mathrm{ph}$ varies as $C_\mathrm{ph}\propto T^3$. In order to improve the reliability at higher temperatures, we use a second term of the harmonic-lattice approximation below 14\,K. The data can be well described by $C_\mathrm{el}+C_\mathrm{ph} = \gamma T + \alpha T^3 + \beta T^5$ (see the dashed line in Fig.\,3(a)), in which the lattice contribution is accounted for by $\alpha T^3 + \beta T^5$. The Sommerfield coefficient $\gamma_n$ values are 5.3(1), 5.1(0.5), 4.9(0.5), and 4.95 for $x$ = 0, 0.04, 0.09, and 0.11 respectively. The estimated universal parameter $\Delta C_{el}/\gamma_nT_c$ of the specific heat at $T_c$ is
$\approx$ 2.14, 2.43, 2.2, and 1.95~mJ/mol K$^2$ for $x$ = 0, 0.04, 0.09, and 0.11, respectively.

According to the BCS theory, the specific heat jump of a superconductor at $T_{c}$ should follow $\Delta C_{p}/\gamma _{n} T_{c}$ = 1.43 in the weak coupling limit. It is so far well reported that a reduced jump in the specific heat $\Delta C_{p}/T_{c}$ compared to that of a single-band s-wave superconductor might be related to a pronounced multiband character of superconductivity with rather different partial densities of states and gap values\cite{M1}. However, jumps of specific heat at $T_c$ in these materials show a deviation from the trend established by Bud$'$ko-Ni-Canfield (BNC) scaling  $\Delta C_{p}/T_{c} \varpropto T^{2}$~\cite{SLB1}. This power law seems to be appropriate for many iron-based superconductors, so far for the 122 systems, i.e., Ba(Fe$_{1-x}$Co$_{x}$)$_{2}$As$_{2}$ and Ba(Fe$_{1-x}$Ni$_{x}$)$_{2}$As$_{2}$,~\cite{SLB2} One of the possible reasons for this universal relation might be a strong pair breaking and the impurity scattering effect in a multiband superconductor~\cite{SLB1,Ko}. Recently, specific heat jump shows also a deviation from that trend in FeSe$_{0.5}$Te$_{0.5}$~\cite{JX}. In addition, the heavily hole-doped superconductors (K,Na)Fe$_{2}$As$_{2}$ and stands out from the other Fe pnictides~\cite{VG}.

From the extracted  $\gamma_n$ values we have estimated the values of universal parameter $C_{el}/\gamma_nT_c$ in Table I, (Fig.\,3(b)). However, a domelike dependence on doping is seen similar to the one in NaFe$_{1-x}$Co$_{x}$As~\cite{Na} and Ba(Fe$_{1-x}$Co$_{x}$)$_{2}$As$_{2}$~\cite{Ba}. The values recorded in in Fig.\,3(b) are larger than the BCS weak-coupling prediction of 1.43. Therefore, we notice that the reduced specific heat jumps are larger than those of optimally doped Ba(Fe$_{1-x}$Co$_{x}$)$_{2}$As$_{2}$~\cite{FH}. Nevertheless, the values are comparable to those of optimally doped Ba$_{1-x}$K$_{x}$Fe$_{2}$As$_{2}$~\cite{PP}. In addition, band-structure calculations in FeSe estimated the value of $\gamma_o$ of about 2.2~mJ/mol K$^{2}$. Since $\gamma_n$ = (1 + $\lambda _{el-bos}$)$\gamma_o$, where $\lambda _{el-bos}$ is the total coupling strength between the quasiparticle (QP) and bosons~\cite{ASu}, the $\lambda _{el-bos}$ value is estimated to be $\approx$ 1.4. Without any model fitting, the values of normalized specific heat jump and the $\lambda _{el-bos}$ constant have already further confirm a stronger electron-boson-coupling strength in FeSe$_{1-x}$S$_{x}$.

\subsection{The upper critical fields $H_{c2}(T)$ and their anisotropy}

\begin{figure}
\includegraphics[width=18pc,clip]{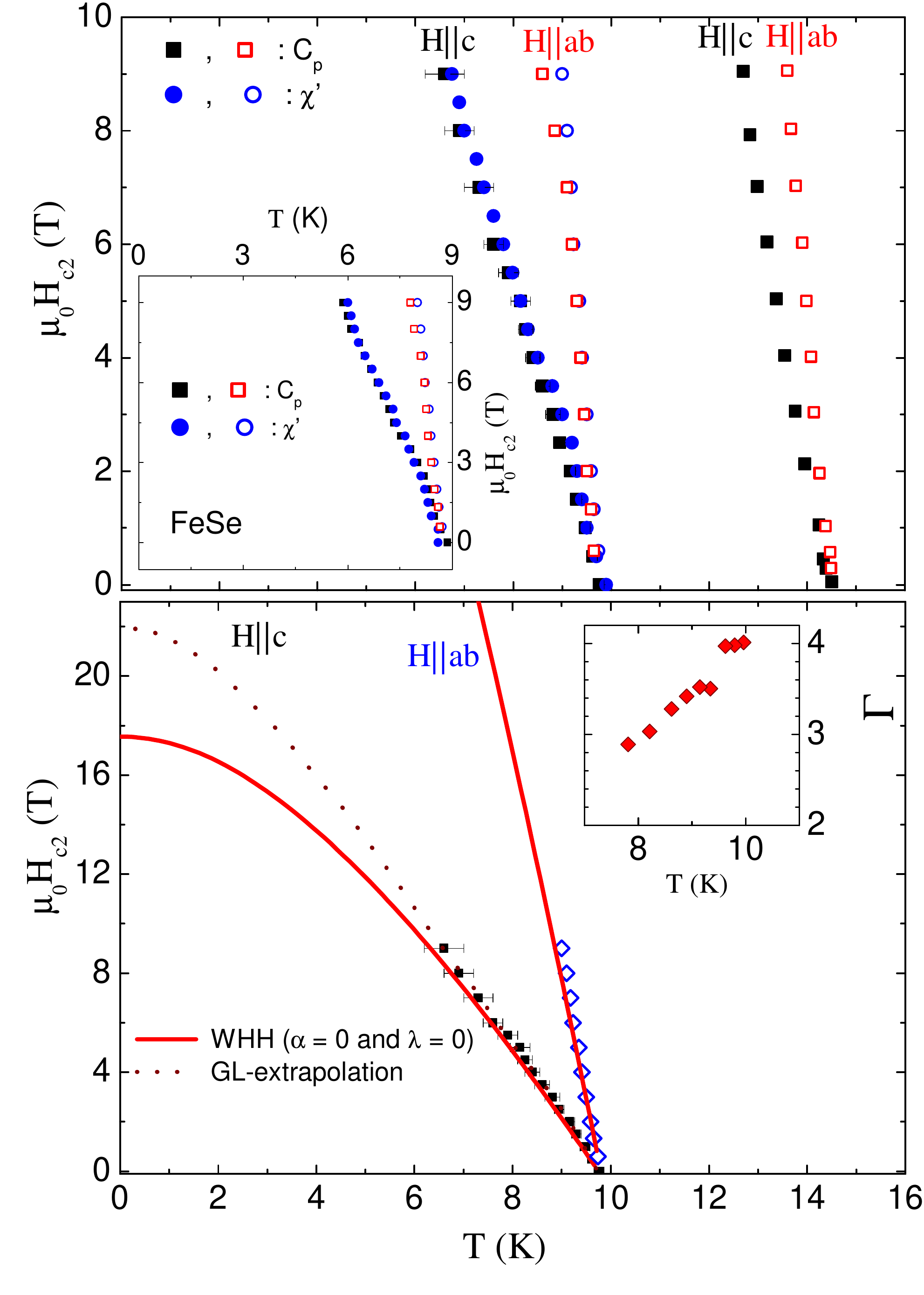}
\caption{Upper panel summarizes phase diagram of $H_{\mathrm{c2}}$ vs.~temperature of  $x$ = 0.04 and FeSe$_{0.45}$Te$_{0.55}$~\cite{Nat} ($T_{c}$ = 14.5\,K) for the field applied parallel and perpendicular to $c$. $T_{c}$ has been estimated from an entropy-conserving construction and AC measurements. $H_{\mathrm{c2}}$ in~\cite{Nat} is determined from transport measurements. The inset illustrates phase diagram of $H_{\mathrm{c2}}$ vs.~temperature of the FeSe. The red solid points in $x$ = 0.04 data are estimated from the ac magnetization while the black data represents the specific heat for $H \parallel c$ and $H \parallel ab$. Lower panel represents the phase diagram with the fit to the Ginzburg-Landau equation (dotted line) and the WHH model for $\lambda = 0$, $\alpha = 0$ (solid lines). The inset shows the anisotropy $\Gamma = H_{\mathrm{c2}}^{B\perp c}/H_{\mathrm{c2}}^{B\parallel c}$ determined by an interpolation of the $H_{\mathrm{c2}}$ curves. The line is a guide to the eye.} 
\end{figure}

Figures 3(c) and (d) summarize the temperature dependence of the specific heat measured at different applied magnetic fields parallel to the $c$-axis (and parallel to the $ab$-plane) as shown in the insets $x$ = 0.04 and 0.09, respectively. In zero-field specific-heat measurements, a very sharp anomaly is clearly seen. This anomaly is attributed to the superconducting transition at $T_{c}$. This specific-heat jump is systematically shifted to lower temperatures upon applying DC magnetic fields of up to 9\,T in both orientations. In order to determine the superconducting transition temperature for each field, an entropy conserving construction has been used. The inset in Fig.\,3(e) presents the temperature dependence of the electronic specific heat near the transition temperature at 0.5\,T$\parallel$ab and 2\,T$\parallel$ab data. The results yield an anisotropy of about $\Gamma$ = 4 for $x$ = 0.04. The extracted data at each field were used to map out the superconducting phase diagram depicting $H_{c2}$ (see Fig.\,4). In order to determine the upper critical field {$H_\textup{c2}$ for the $c$} orientation, we use the Ginzburg Landau (GL) equation as an initial step:~\cite{Woollam1974}

\begin{equation}\label{eq1}
H_{c2}=H_{c2}(0)[\frac{1-t^{2}}{1+t^{2}}],
\end{equation}
where $t$ is the reduced temperature $t = T/T_{c}$. The fit is shown with a dotted line in the inset of Fig.\,3(f). Another possibility to extract the upper critical field $H_{c2}(0)$ is to consider the single-band-Werthamer-Helfand-Hohenberg (WHH) formula~\cite{NR} for an isotropic one-band BCS superconductor in a dirty limit. An example of WHH fit is shown with the solid lines in the inset of Fig.\,3(f) for both orientations. The WHH theory ($\alpha = 0$, $\lambda _{so} = 0$) predicts the behavior of $H_\mathrm{c2}(T_\mathrm{c})$ taking into account paramagnetic and orbital pair-breaking\cite{NR}. Here, $\alpha$ is the Maki parameter which describes the relative strength of orbital breaking and the limit of paramagnetism. $\lambda$$_{so}$ (when $\lambda$$_{so}$ $>$ 0, the effect of the spin-paramagnetic term) is the spin-orbit coupling strength. The orbital limiting field $H_{\mathrm{c2}}^{\mathrm{orb}}$ at zero temperature is determined by by a slope at $T_\mathrm{c}$ as $\mu_{0}H_{\mathrm{c2}}^{\mathrm{orb}} = 0.69 \, T_{\mathrm{c}}\, (\partial \mu_{0}H_{\mathrm{c2}}/\partial T)|_{T_{\mathrm{c}}}$, where $\frac{d\mu_{0}H_{c2}}{dT}|_{Tc}$ is indicated by the slopes of the fitted straight lines. It should be borne in mind that Fig.\,4 (upper panel) reports in the same plot the data for FeSe$_{0.45}$Te$_{0.55}$~\cite{Nat} and the anisotropy of $H_{c2}(T)$ is found to be 2.

The upper critical field values at $T$ = 0 for the FeSe and FeSe$_{0.96}$S$_{0.04}$ have been evaluated to be $\mu_0H_{c2}^{(c)}(0)$ = 12.8(1), 17.5(1)\,T and $\mu_0H_{c2}^{(ab)}(0)$ = 24.4(2), 67.5(2)\,T. The anisotropy is found to be $\Gamma = H^{ab} _{c2}/H^{2} _{c2} \approx$~2 and 4 for $x$ = 0 and 0.04, respectively. The extracted values for $H_{c2}(0)$ are summarized in Table I for the other samples. The observed small differences between the data obtained from the specific heat and the AC magnetization for $H \parallel c$ (see upper panel Fig.\,4) is not surprising since these methods naturally imply different criteria for the $T_{c}$ determination. It is evident that, the one-band WHH model fails to satisfy the extracted $H_{c2}(0)$, \emph{i.e.,} the fit shown by the red solid line in the lower panel of Fig.\,4. Therefore, we believe that the observed deviation from the single band WHH model is related to multiband effects. Furthermore, the temperature dependence in $H_{c2}(T)$ displays a non-linear behavior and shows a slightly concave curvature at low temperatures. This behavior is reminiscent of the one reported in Fe-based superconductors~\cite{M1,SK} in which similar $H_{c2}(T)$ curves were well described by a multiband effect. This claim is supported by the indications of a two-band-like behavior from the zero-field specific heat measurements and the London penetration depth (discussed below). The calculated upper critical field $H_\textup{c2}(0)$ and average slope
$-dH_\textup{c2}^\textup{(c)}/dT$ values of the superconducting single-crystal samples are compared in Table I.

In all of the investigated S-doped samples, both specific heat and AC susceptibility measurements at 9\,T$\parallel ab$ data show a sharper transition compared to the 9\,T data $\parallel c$, indicating a highly anisotropic crystal, $[$see Figs.\,3(e) and (f)$]$. From the behavior of $H_\mathrm{c2}$ vs.~$T$ for the different field orientations we calculate the anisotropy $\Gamma = H_{\mathrm{c2}}^{B\perp c}/H_{\mathrm{c2}}^{B\parallel c}$ using a linear interpolation. The results are outlined in the inset of the lower panel of Fig.\,4. Our data allow tracking of $\Gamma$ up to temperatures very close to $T_{c}$. $\Gamma$ increases upon approaching $T_\mathrm{c}$ and reaches a value of about 4 at the critical temperature for $x$ = 0.04. This finding is in contrast to the results found in other Fe-based superconductors, which suggests a saturation or even a decrease of $\Gamma$ close to $T_\mathrm{c}$~\cite{Lee}. This indicates that an orbital pair breaking is dominating the suppression of superconductivity close to $T_c$. At lower temperatures, the isotropic Zeeman pair breaking becomes more important, which leads to lowering of $\Gamma$. Moreover, the strong $T$-dependent superconducting anisotropy $\Gamma = H_{\mathrm{c2}}^{B\perp c}/H_{\mathrm{c2}}^{B\parallel c}$ provides further evidence for multiband scenario as in the case of the two-band superconductor MgB$_{2}$~\cite{A1}. Surprisingly, this anisotropy is considerably larger than the typical value of $\Gamma \sim$ 2 and 2.6 found in nearly optimally hole-doped (BaK)Fe$_{2}$As$_{2}$ \cite{U2009} and in FeSe$_{0.45}$Te$_{0.55}$~\cite{Nat} ($T_{c}$ = 14.5\,K), but lower than the ones determined in SmFeAsO$_{0.85}$F$_{0.15}$ and La(O,F)FeAs thin films~\cite{U2011,Backen2008}. On the other hand, these values are comparable with $\Gamma$ values of e.g., KFe$_{2}$As$_{2}$ \cite{M1} and LaFePO \cite{JJ}.

\subsection{Superconducting energy-gap structure}

\subsubsection{London penetration depth}

\begin{figure}[tbp]
\includegraphics[width=18pc,clip]{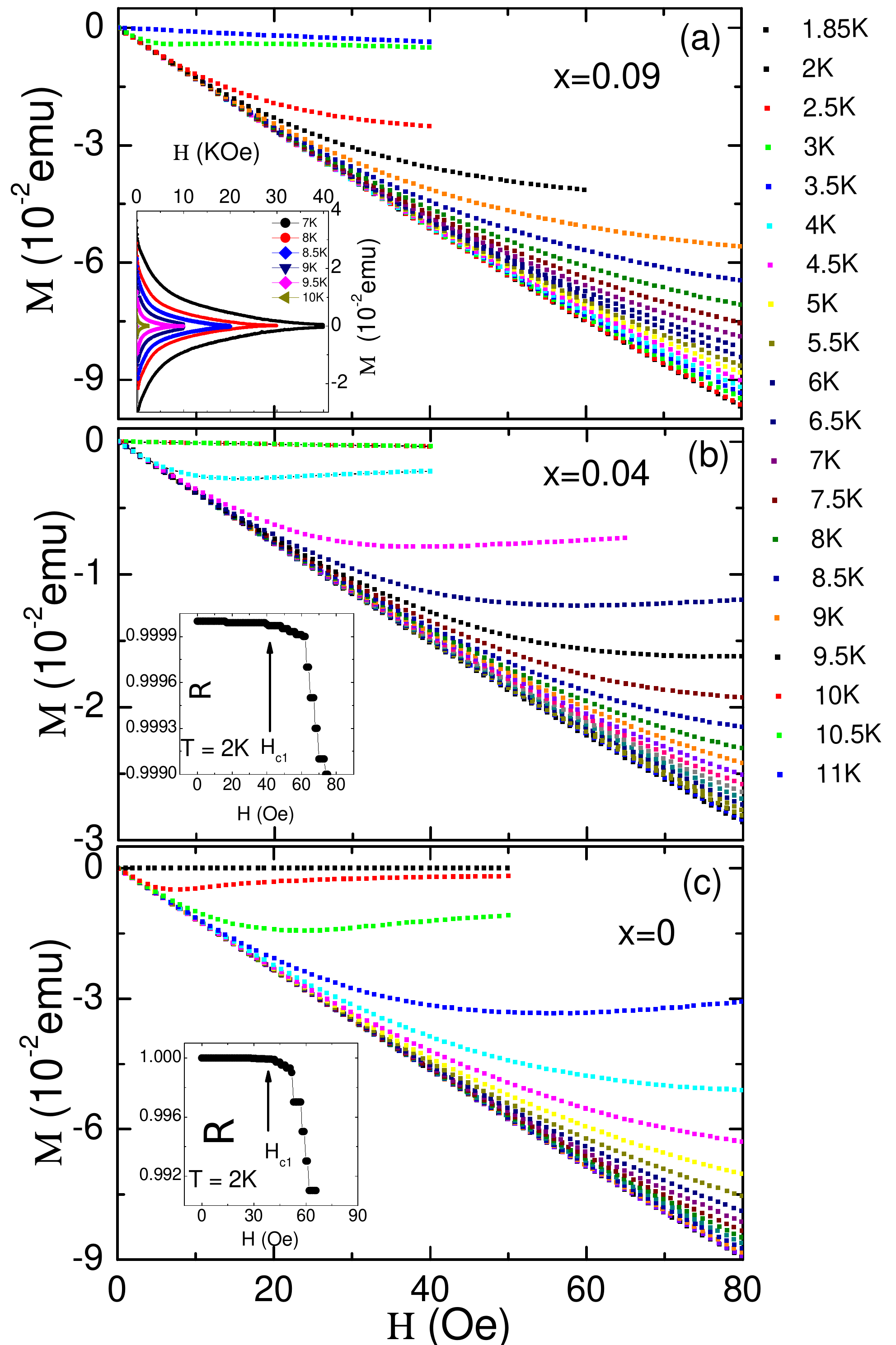}
\caption{The field dependence of the initial diamagnetic part of the magnetization curves measured at various temperatures for $H \parallel c$ in FeSe$_{1-x}$S$_{x}$ single crystal, where $x$ = 0, 0.04, 0.09, respectively. The inset of (a) presents the magnetic field dependence of magnetization in FeSe$_{0.91}$S$_{0.09}$ single crystal at different temperatures close to $T_{c}$. The insets in (b) and (c) depict an example used to determine the $H _{c1}$ value using the regression factor $R$, at $T$ = 2\,K.}
\end{figure}

\begin{figure*}
\includegraphics[width=30pc,clip]{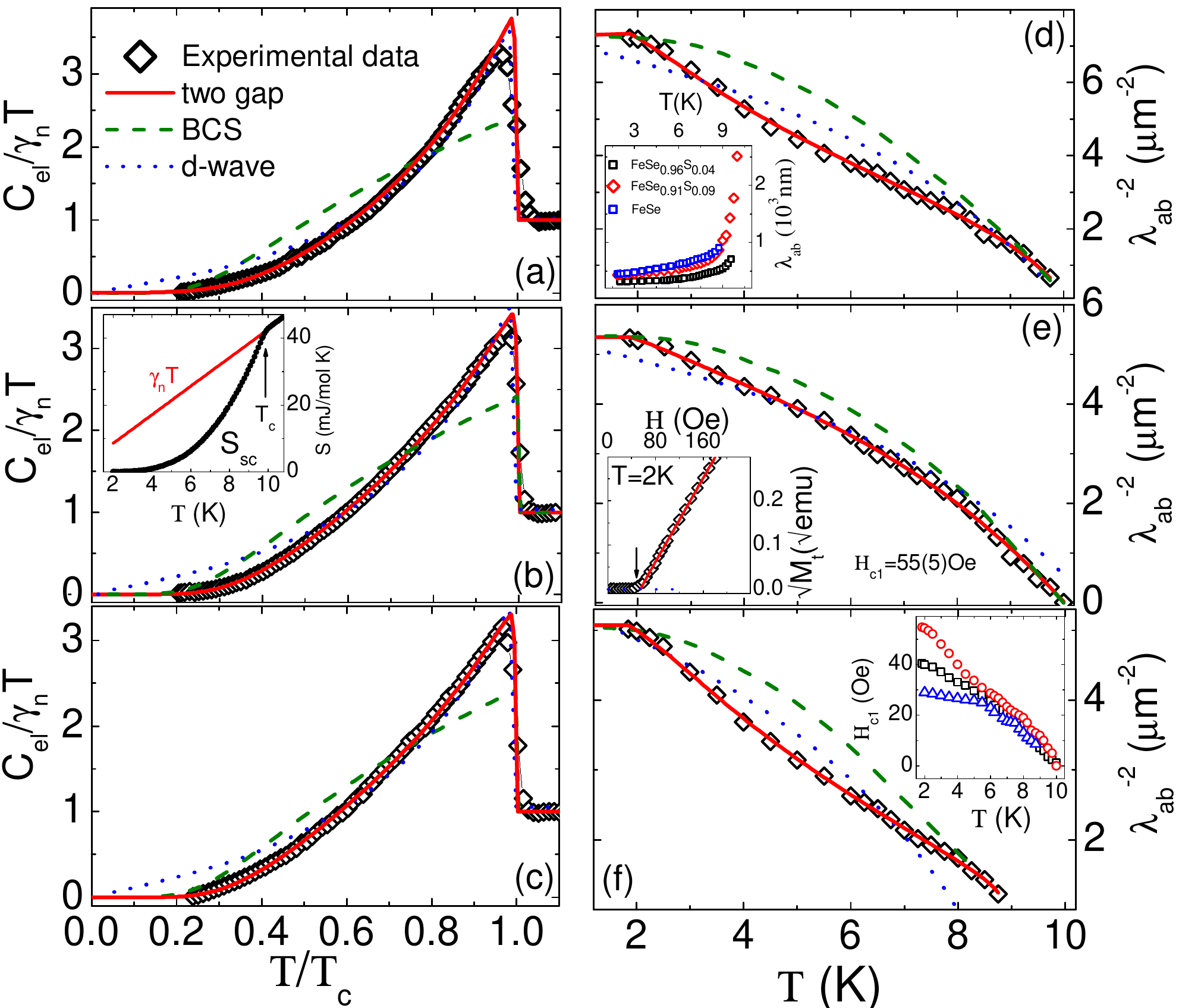}
\caption{\label{fig:wide} (a-c) The normalized SC electronic specific heat of the three samples after subtracting the phonon contribution as a function of reduced temperature , {$T/T_\textup{c}$}. The inset in (b) presents the entropy in the normal and SC state as a function of $T$. (d-f): The temperature dependence of the London penetration depth for $x$ = 0, 0.04, and 0.09, respectively. The inset of (d) presents the temperature dependence of the magnetic penetration depths $\lambda _{ab}$ vs. $T$. The inset of (e) depicts an example used to determine the $H _{c1}$ value using the trapped moment at $T$ = 2\,K of the typical plot of $\sqrt{M_{t}}$ vs. $H$. The solid red line is a linear fit to the high-field data of $\sqrt{M_{t}}$ vs. $H$. $H_{c1}$ values are determined by extrapolating the linear fit to $\sqrt{M_{t}}$ =0. The inset in (f) represents the phase diagram of $H_{\mathrm{c1}}$ for the field applied along the $c$ axis. The dashed lines represent the theoretical curves based on single-band weak-coupling BCS theory, while the dotted lines present the $d$-wave approximation. The solid lines represent the curves of the two $s$-wave gap model.}
\end{figure*}

The London penetration depth $\lambda$, is a fundamental parameter characterizing the SC condensate and probes the gap structure of bulk superconductors. The $T$-dependence of $\lambda$ is directly determined by the gap function $\Delta(T)$. $\lambda(T)$ = $\lambda(T=0)$+$\delta\lambda(T)$ behaves as $\delta\lambda(T) \propto\exp(\frac{-\Delta}{\kappa _{B}T})$ at low $T$ reflecting a nodelsss superconducting gap $\Delta$ with s-wave symmetry. In $d$-wave pairing scenario containing line nodes,
$\delta\lambda(T) \propto T$ at low $T$ in the clean limit. The experimental determination of the London penetration depth $\lambda(0)$ via measurement of the lower critical field $H_{c1}$ is challenging since $H_{c1}$ is an equilibrium thermodynamic field. The temperature dependence of the SC penetration depth provides another method for detecting the existence of multiple gaps~\cite{Pro}. A popular approach to measuring $H_{c1}$ is by tracking the virgin $M(H)$ curve at low fields and identifying the deviation from the linear Meissner response which would correspond to the first vortex penetration (see Fig.\,5). This technique implicitly relies on the assumption that no surface barriers are present. We have confirmed the absence of the surface barriers in our case from the very symmetric DC magnetization hysteresis curves M(H) (see Fig.\,1(b) and (c) and the inset in Fig.\,5(a)). On the other hand, if surface barriers were predominant, the first vortex entrance would occur at much higher field ($\sim H_c$). Thus absence of surface barriers is a very important for obtaining reliable estimates of the thermodynamic lower critical field. The transition from linear to non-linear $M(H)$, was determined by a user-independent procedure consisting of calculating the regression coefficient $R$ of a linear fit to the data points collected between $0$ and $H$, as a function of $H$ (see the insets in Fig.\,5). In contrast to tracking the virgin $M(H)$ curve at low fields at several temperatures, in which case a heavy data post-processing is needed, (see Fig.\,5), here a careful measurement protocol needs to be followed with little data analysis. Indeed, the $H_{c1}$ values from the virgin magnetization data are close to those obtained from the onset of the trapped flux moment $M_{t}$ $[$see the inset in Fig.\,6(e)$]$. Here, $M_{t}$ is obtained by following sequence: (i) warming the sample up to temperatures above $T_{c}$ i.e., 12\,K, then (ii) cooling the sample in zero field down to the particular $T$, and, subsequently (iii) increasing the applied magnetic field to a ceratin maximum value $H_{m}$ and in a last step (iv) measuring the remanent magnetization $M_{t}$ after the applied field has been removed. The field $H_{m}$ at which $M_{t}$ deviates from zero determines the $H_{c1}$ value at the desired temperature. Then, the extrapolation $\sqrt{M_{t}}\rightarrow 0$ determines the exact value of the $H _{c1}$. The inset of Fig.\,6(e) presents the typical plot of $\sqrt{M_{t}}$ vs. the applied field $H$, for FeSe$_{0.96}$S$_{0.04}$ single crystals. The solid line is a linear fit to the high-field data of $\sqrt{M_{t}}$ vs. $H$. $H_{c1}$ is determined by intercept of the fit with the abscissa.

\begin{table*}
\caption{\label{tab:table 1}  Compilation of the superconducting parameters of samples with various $T_{c}$. We show the {$T_{c}$ (K), $J_{c}$ (10$^{4}$\,A/cm$^{2}$), $\lambda _{ab}(0)$ ((15)\,nm), $\Gamma =
H_{\mathrm{c2}}^{B\perp c}/H_{\mathrm{c2}}^{B\parallel c}$, $\gamma_{n}$ = (mJ/mol K$^{2}$), the universal parameter ($\Delta C_{el}/\gamma _{n}T_{c}$), $-\frac{d\mu_{0}H_{c2}}{dT}|_{Tc}$ (T/K), upper critical field $H_{c2}$ (T), $\beta$ = 10$^{-4}$\,mJ/mol K$^{4}$, $\alpha$ = 10$^{-7}$\,mJ/mol K$^{6}$, $d$-wave ($\Delta_{0}/k_{B}T_{c}$), the superconducting gap ratio ($\gamma_{1}, \gamma_{2}/\gamma_{n}$), and two $s$-wave gaps ($\alpha_{1} = \Delta_{1}/k_{B}T_{c}$ and $\alpha_{2} = \Delta_{2}/k_{B}T_{c}$) extracted for the investigated samples.}}
\begin{ruledtabular}
\begin{tabular}{cccccccccccccc}
$x$ &$T_{c}$ &$J_{c}$ & $\lambda _{ab}(0)$& $\Gamma$ & $\gamma_{n}$ & $\Delta C_{el}/\gamma _{n}T_{c}$ & $-\frac{d\mu_{0}H_{c2}}{dT}|_{Tc}$& $H_{c2}$ &  $\beta$ & $\alpha$ & $d$-wave & $\gamma_{1}, \gamma_{2}/\gamma_{n}$ & $\alpha_{1}  /  \alpha_{2}$ (C$_{p}$, $\lambda _{ab}$)\\
\hline
\\
0 &8.5 &1.1&446&2 &5.3 & 2.14 &2.1&12.8 &4.34&-0.384&2.8&0.4, 0.6 & 0.88, 0.79 / 2.22, 2.05\\

\\
0.04 &9.58&1.3&372&4 &5.1 & 2.43 &2.6&17.5 &4.8&-3.62&2.36&0.44, 0.56 & 1.9, 1.85 / 2.5, 2.3\\

\\
0.09 &10.1&1.35&433&3.5&4.9 & 2.2 &2.7&19 &3.6&-2.5&3.05&0.47, 0.53 & 2.2, 2.1 / 2.35, 2.28\\

\\
0.11 &10.7 & 1.45&415&3.5&4.95 & 1.95&2.73&20.2 &4.1&-2.9&3.12& 0.42, 0.58 & 1.96 / 2.39\\
\end{tabular}
\end{ruledtabular}
\end{table*}

The above measured values of $H_{c1}$ need to be corrected due to the finite demagnetization effects. Indeed, the deflection of field lines around the sample leads to a more pronounced Meissner slope given by $M/H_{a} = -1/(1-N)$, where $N$ is the demagnetization factor. Taking into account these effects, the absolute value of $H _{c1}$ can be estimated by using the relation proposed by Brandt~\cite{Brandt}. For our sample we find $N$ $\approx0.96$, 0.95, 0.97, and 0.96 for FeSe$_{0.96}$S$_{0.04}$, FeSe$_{0.91}$S$_{0.09}$, FeSe$_{0.89}$S$_{0.11}$ and FeSe respectively. In order to shed light on the pairing symmetry in our system, we estimated the penetration depth using the traditional Ginzburg-Landau (GL) theory, where $H_{c1}$ is given by: $\mu_{0}H_{c1}^{\parallel c} = (\phi$$_{0}/4\pi\lambda _{ab}^{2})\ln\kappa _{c}$, where $\phi$$_{0}$ is the magnetic-flux quantum $\phi$$_{0}$ = $h/e^{\ast}$ = 2.07 x 10$^{-7}$Oe cm$^{2}$, $\kappa _{c}$ =$\lambda _{ab}$/$\xi _{ab}$ is the Ginzburg-Landau parameter. The value of $\kappa$ was determined from the equation:$\frac{2H_{c1}(0)}{H_{c2}(0)} = \frac{\ln\kappa+0.5}{\kappa^{2}}$. It should be mentioned that the SC penetration depth is a very important physical quantity and it is sensitive to the absolute value the order parameter(s); and, in that sense, also sensitive to any nodes or a deep local minima of the gap. It is worth mentioning that the SC penetration depth is also dependent on the distribution of Fermi velocities. In this context it is most sensitive to fast electrons in sharp contrast to the upper critical field which is highly sensitive to the subgroup of electrons with low Fermi velocities.

In Fig.\,6(d-f), we analyze the temperature dependence of the London penetration depth for the samples with $x$ = 0, 0.04, and 0.09, respectively. We compare our data to the $d$-wave and single-gap BCS theory under the weak-coupling approach (see dotted and dashed lines in Fig.\,6(d-f)). Indeed, both quantities lead to a rather different trend and show a systematic deviation from the data in the whole $T$-range below $T_{c}$. On the other hand, we also apply a phenomenological two-gap model which is in line with the multigap-superconductivity reported by Carrington and Manzano~\cite{Carrington}. Within this model, the temperature dependence of each energy gap can be approximated as:~\cite{Carrington,Car1,Car2}  $\Delta _{i}(T) = \Delta _{i}(0) {\tanh[1.82(1.018(\frac{T_{ci}}{T}-1))^{0.51}]}$. According to Ref.~\onlinecite{VAG}, for each band, $\lambda _{i}^{-2}(T)$  is given by:

\begin{equation}
\label{eq2} \lambda _{i}^{-2}(T) = \frac{\Delta _{i}(T){\tanh(\frac{\Delta _{i}(T)}{2k_{B}T})}}{\lambda _{i}^{2}(0)\Delta _{i}(0)},
\end{equation}
where $\lambda _{i}(0)$ is the residual penetration depth for each band, $k_{B}$ is the Boltzmann constant. Considering different partial contributions of each band to the overall $\lambda(T)$, we use the following expression: $\lambda^{-2}(T) = r\lambda _{1}^{-2}(T) + (1-r)\lambda _{2}^{-2}(T)$ with $r$ being the weighting factor indicating the contribution of the small gap. The best description of the experimental data is obtained using values of $\Delta_{1}/k_{B}T_{c}$ = 1.72$\pm0.3$, 1.79$\pm0.25$, and 0.79$\pm0.15$ and $\Delta_{2}/k_{B}T_{c}$ = 2.28$\pm0.3$, 2.1$\pm0.25$, and 1.95$\pm0.2$ for $x$ = 0.09, 0.04, and 0, respectively. The weighting factor is found to be around $r$ = 0.25$\pm0.08$, 0.38$\pm0.1$, and 0.22$\pm0.2$  for $x$ = 0, 0.04, and 0.09, respectively. The fits are represented by solid red lines in Fig.\,6(d-f). The extracted gap values for FeSe are comparable to those obtained from the two-band $s$-wave fit of the specific heat data and the Andreev reflection spectroscopy results~\cite{JYLin,DC}. It is worth pointing out that the $\lambda^{-2}_{ab}(T)$ of the SC samples $[$see the inset of Fig.\,6(d)$]$ does not saturate at low temperatures, as it could be expected for a fully gapped clean $s$-wave superconductor. $\lambda _{ab}(T)$ is nearly constant at low temperatures, which demonstrates negligible quasiparticle excitations. The above penetration depth results are consistent with the presence of two $s$-wave-like gaps. Both gap values in the S-doped samples are considerably larger than the BCS weak-coupling limit. These observations show clearly that there are no nodes in the SC energy gap indicating a strong-coupling multiband (and nodeless) superconductivity in iron chalcogenide Fe(Se,S) superconductors. A similar possible strong coupling multiband superconductivity in Fe(Se,Te) has been conjectured from a detailed penetration depth and specific heat experiments~\cite{HKi,JHu}. {The temperature dependence of the magnetic penetration depth of $d$-wave superconducting gap calculations was performed by using the following functional form:~\cite{RKH,RKH2}}
\begin{equation}
\label{eq2} \frac{\lambda _{ab}^{-2}(T)}{\lambda _{ab}^{-2}(0)} = 1 + \frac{1}{\pi}\int_{0}^{2\pi} \int_{\Delta(T,\varphi)}^{\infty} (\frac{\partial f}{\partial E})\frac{EdEd\varphi}{\sqrt{E^{2}-\Delta(T,\varphi)^{2}}},
\end{equation}
{where $f = [1 + \exp(E/K_{B}T)]^{-1}$ is the Fermi function, $\varphi$ is the angle a long the fermi surface, and $\Delta(T,\varphi)$ = $\Delta_{0}\delta(T/T_{c})g(\varphi)$ ($\Delta_{0}$ is the maximum gap value at $T$=0). The function $g(\varphi)$ is given by $g^{d}(\varphi)$ = $|cos(2\varphi)|$ for the $d$-wave gap. The results of the analysis are presented in Fig.\,6(d-f) by dotted lines. The fit to of the
experimental data for the $d$-wave case we get for $\Delta_{0}/k_{B}T_{c}$ = 2.8, 2.36, and 3.05 for $x$ = 0, 0.04, and 0.09. It is obvious that the $d$-wave case cannot describe the penetration depth data. On the other hand, the experimental data are well described by the two-gap $s$ wave models.

It is noteworthy that in a FeTe$_{0.58}$Se$_{0.42}$ system, a careful analysis of the superconducting and normal state properties indicates a possibility of strong coupling superconductivity~\cite{K1}. This study is followed by precise measurements of the temperature dependence of the London penetration depth by Cho {\it et al.} Their analysis strongly suggest a presence of two $s$-wave-like gaps with magnitudes $\Delta_{1}/k_{B}T_{c} = 1.93$ and $\Delta_{2}/k_{B}T_{c} = 0.9$~\cite{C1}. These two precise measurements~\cite{K1,C1} were followed by the comment of Klein {\it et al.}~\cite{K2} and response of K. Cho {\it et al.}~\cite{C2}. In the latter case, the authors have shown convincingly that previous studies~\cite{HKi} most likely have issues with pair-breaking scattering. The authors have reported that the presence of strong scattering hinders any determination of gap values from the temperature dependence of the superfluid density.

\subsubsection{Specific heat}

The normalized zero-field data $C_\textup{el}$/$\gamma_\textup{n}$\emph{T} as a function of the reduced temperature {$T/T_\textup{c}$}, obtained after subtracting the $C_{ph}$, is presented in Fig.\,6(a-c) together with the fits to various models. It is obvious from Fig.\,6(a-c) that the superconducting transition at $T_\textup{c}$ is well pronounced, with a sharp jump in $C_\mathrm{el}$ at $T_c$. The entropy conservation required for a second-order phase transition is fulfilled as shown in Fig.\,6(b). This check warrants the thermodynamic consistency for both: the measured data and the determination of $C_\mathrm{el}$. We have attempted best fits to the data using three different models: single-band weak-coupling BCS theory with the $s$-wave gap $\Delta_{(0)}/k_{B}T_{c}$ = 1.76; a $d$-wave calculation using $\Delta=\Delta_{(0)}\cos(2\phi)$; and two-gaps $s$-wave in Figs.\,6(a-c). Below $T_{c}$ we observe systematic deviation of both single-gap and the $d$-wave fit from the data showing a higher jump at $T_{c}$ than the $s$-wave model. Thus we focus our discussion on the possibility of two SC energy gaps using the generalized $\alpha$-model, that explains the specific heat behavior in a multiband superconductors~\cite{Bouquet2001}. The corresponding fits are shown in Figs.\,6(a-c). Although the two-gap model contains two distinct gaps, the specific heat value is calculated as the sum of contributions, each one following the BCS-type temperature dependence, $\Delta(0) = \gamma_{1}\Delta_{1}(0) + \gamma_{2}\Delta_{2}(0)$~\cite{Bouquet2001} and the thermodynamic properties are obtained as the sum of the contributions from the individual bands, i.e., $\alpha_{1} = \Delta_{1}/k_{B}T_{c}$ and $\alpha_{2} = \Delta_{2}/k_{B}T_{c}$.

The estimated $\Delta _{1}(0)/k_{B}T_{c}$ for the small gap for $x$ = 0, 0.04, and 0.09 is 0.88$\pm0.1$, 1.9$\pm0.2$, and 2.2$\pm0.2$, while the large gap $\Delta _{2}(0)/k_{B}T_{c}$ is found to be 2.2$\pm0.2$, 2.5$\pm0.2$, and 2.35$\pm0.2$, for $x$ = 0, 0.04, and 0.09 respectively. The calculated data and the relative weights are illustrated in red lines in Fig.\,6(a-c). The error bars represents the width of the corresponding range of gap amplitudes obtained in the fit for both values of $\Delta _{1}(0)/k_{B}T_{c}$ and $\Delta _{2}(0)/k_{B}T_{c}$. The results obtained in the present work are consistent with ones of the models considered in Ref.~\cite{JYLin}. The ratio of the two gaps ($\Delta _{1}(0)/\Delta _{2}(0)$, is $\approx$ 0.7 and 0.9 for $x$ = 0.04, and 0.09, respectively) is comparable to the FeSe$_{0.43}$Te$_{0.57}$ case and it is noticeably larger than in iron pnictide superconductors (between 0.24 and 0.5)~\cite{FH,PP}. All of the fitting parameters are remarkably consistent with those obtained from the penetration depth measurements. They give a strong evidence for a two-gap SC at a Fe(Se,S) system. It has been theoretically demonstrated that in multiband superconductors if the ratio of two isotropic $s$-wave gaps $\Delta _{1}(0)/\Delta _{2}(0) > 0.5$, the field-induced low energy excitations would be less pronounced compared to a single-band $s$-wave symmetry~\cite{Bang}. With this respect, in the low field range $\gamma(H)$ would slowly increase with $H $ (see Fig.\,7). As mentioned above, our obtained ratio of the two gaps is higher compared to the critical value of 0.5 suggested by the theory, which further confirms the multiband nature in Fe(Se,S). In addition, the obtained two gaps of both S-doped samples are consistent with the penetration depth results and larger than the BCS value in the weak-coupling regime. Overall, such a behavior confirms the strong coupling nodeless superconductivity in Fe(Se,S).

\begin{figure}
\includegraphics[width=20pc,clip]{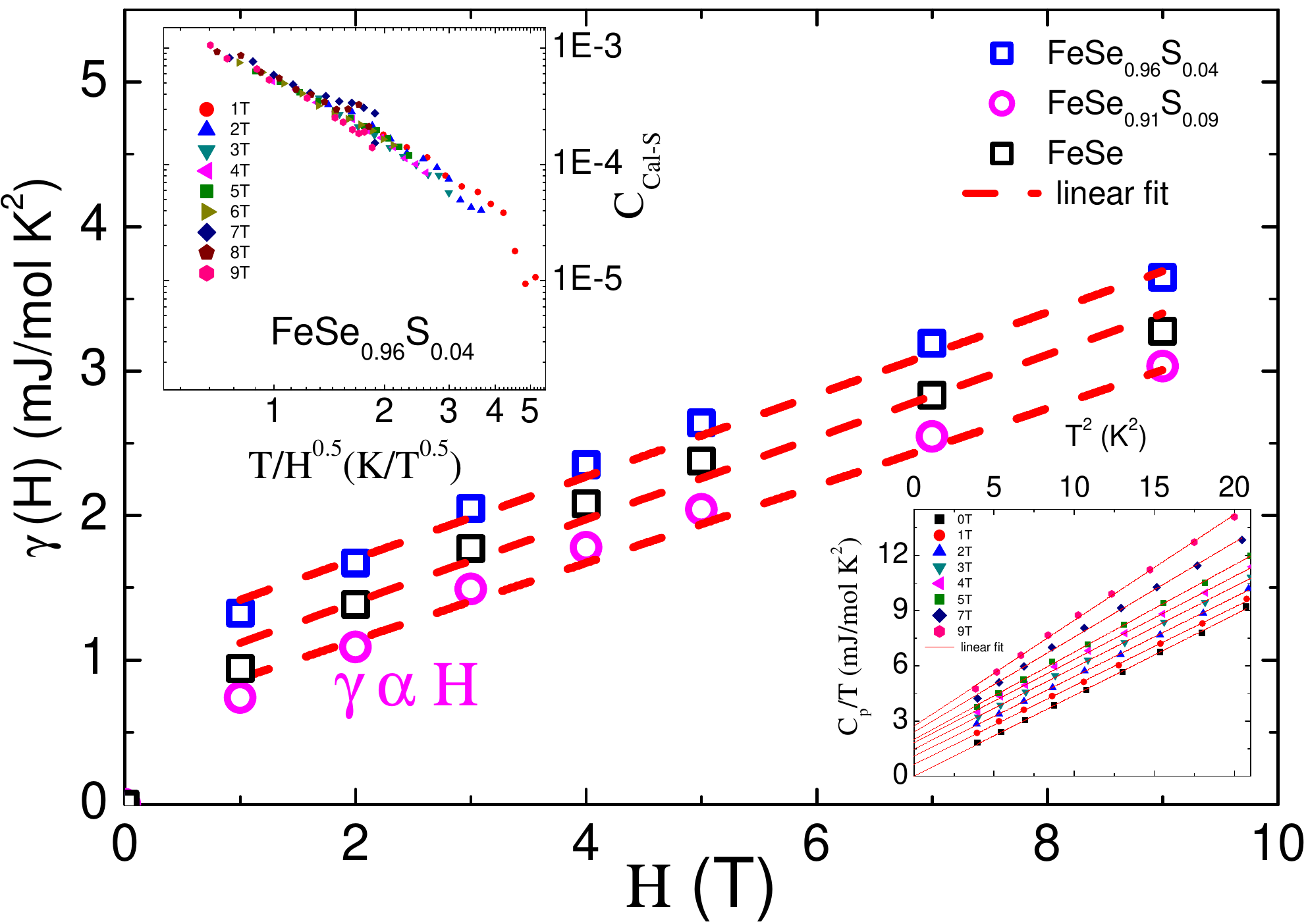}
\caption{The field dependence of the mixed state quasiparticle contribution $\gamma(H)$ for $H\parallel$c for $x$=0, 0.04, and 0.09. The dashed lines represent the phenomenological linear fits above $H$ = 1\,T. The upper inset presents the scaling of the data according to the $s$-wave scenario: $C_{cal-s}$ = [C(H) - C(0)]/$T^{3}$ vs. $T/\sqrt{H}$. The lower inset shows the specific heat of FeSe$_{0.96}$S$_{0.04}$ plotted as $C_{p}/T$ vs. $T^{2}$ measured under various magnetic fields up to 9\,T in the low temperature region. The solid lines show a linear extrapolation of the data.} 
\end{figure}

Next we discuss the field dependence of specific heat, which is another independent, sensitive test of the gap structure. It has been well demonstrated that in the case of an isotropic $s$-wave superconductor, $\gamma(H) \propto H$ because the specific heat in the vortex state is dominated by the contribution from the localized QP in the vortex core. Recently, Storey {\it et al.}~\cite{JGS} pointed out that the number of Caroli-de Genned bound states increases linearly with the field due to the linear increase in the number of vortices entering the sample. On the other hand, for the line nodes $\gamma(H) \propto H^{0.5}$, the QPs contributing to the density-of-states (DOS) come from regions away from the vortex core and close to the nodes and supercurrents around the vortex
core in the mixed state causing a Doppler shift of the QP excitation spectrum~\cite{GEV}. The temperature dependence of the low-$T$ part of the specific heat data measured in various magnetic fields applied along the \emph{c} axis is shown in the lower inset of Fig.\,7. The data plotted as $C_p/T$ vs. $T^{2}$ fits to $C_{p}$/T = $\gamma_n$ + $\beta T^{2}$, with $\gamma_n$ and $\beta$ as electronic and lattice coefficients, respectively. It should be mentioned that the absence of the so-called $\gamma_r$ at the linear-$T$ term of the zero-field specific heat indicates high quality of the single crystals. Nearly perfect linear behavior without any magnetic impurities have been observed  in our samples (see Fig.\,7). The applied magnetic field enhances the low-$T$ specific heat, indicating the increase of the QP DOS at the Fermi level induced by a magnetic field. A linear extrapolation of the low-$T$ data to zero temperature yields the field dependence of the field-induced contribution. The main panel of Fig.\,7 presents the field dependence of the specific heat coefficient. The dashed lines are linear fits for $H \parallel c$ above $H$  = 1\,T as anticipated for a case of nodeless SC gap.

Further confirmation of the nodeless character of superconductivity in our investigated systems comes from the low temperature specific heat data of the finite-temperature region in the mixed state. In fact, the quasiparticle excitations in superconductors with different gap symmetries are obviously distinct. In $s$-wave superconductors, the inner-core states dominate quasiparticle excitations and a simple scaling law proposed by Liu \emph{et al}., holds in the case of possible $s$-wave gap in Sr$_{0.9}$La$_{0.1}$CuO$_{4}$~\cite{Liu2005}:

\begin{equation}\label{eq1}
C_\textup{QP}/T^{3} \approx C_\textup{core}/T^{3} = \gamma_\textup{n}/H_\textup{c2}(0) \times (T/\sqrt{H})^{-2}
\end{equation}
where $C_\textup{QP}$ and $C_\textup{core}$ are the specific heat of quasiparticles induced by the applied magnetic field and quasiparticles present inside from the Abrikosov vortex cores in the mixed state, respectively. The $s$-wave scenario of the scaling result of the field-induced term in the mixed state is presented in the upper inset of Fig.\,7. All the data at different magnetic fields can be roughly scaled within the $s$-wave scenario in one line. Similar low temperature specific heat studies have been already conducted on several Fe-based superconductors~\cite{Zeng2011,GMu2009}.

For the sake of comparison, in Table I we have summarized the superconducting parameters for the investigated samples extracted from this study. According to Fig.\,1(d) and Table I, with increasing S content the critical current density, which is a measure of the strength of the pinning force density and can be very conveniently used to characterize the strength of disorder in the system, enhances, suggesting improved flux pinning in those samples. The absolute value of the penetration depth in the $T\rightarrow0$ limit determined for FeSe$_{0.96}$S$_{0.04}$, FeSe$_{0.91}$S$_{0.09}$, and FeSe, yields $\lambda _{ab}(0)$ = 372(15), 433(15), and 446(15)\,nm, respectively. These values are somewhat smaller than 560(20)\,nm found in Fe(Te,Se)~\cite{HKi}, but comparable to the FeSe$_{0.85}$ and FeSe~\cite{RKH,HA}. In Fig.\,6(d-f), a kink structure is observed on the $\lambda^{-2}_{ab}(T)$ curves. This kink in $\lambda^{-2}_{ab}(T)$ can be associated with the two-band supercondictivity as in the cases of Fe(Te,Se), Ba$_{0.6}$ $_{0.4}$Fe$_{2}$As$_{2}$, and MgB$_{2}$~\cite{HKi,CR,Mg1}. The upper critical field, $H_{c2}$, for the S-doped FeSe sample increases with increased doping, which is mainly due to its enhanced $T_{c}$. From $\gamma_n$ values, we estimate the universal parameter $\Delta C_{el}/\gamma_n T_{c}$, {which is} considerably higher than the prediction of the weak coupling BCS theory ($\Delta C_{el}/\gamma_n T_{c}$ = 1.43). Taking into account the fact that the superconducting transition is relatively sharp $[$see Fig.\,6(a-c)$]$, a distribution in $T_\textup{c}$ or the presence of impurity phases {cannot} explain the higher value of the universal parameter. We believe that the presence of strong coupling superconductivity explain this higher values. Most remarkably, the specific-heat data allows for precise evaluation of SC volume fraction (V$_{SC}$), i.e., V$_{SC}$ = ($\gamma _{n} - \gamma _{r}$)/$\gamma _{n}$, with $\gamma _{r}$ being the residual electronic specific-heat coefficient. Since our $\gamma _{r}$ is almost absent (see lower inset in Fig.\,7), V$_{SC}$ estimated from specific heat is in fair agreement with our magnetization data $[$see Fig.\,1(a)$]$. The overall values of the investigated superconducting gap derived from specific heat is similar to the one obtained from the penetration depth. However, both large gap, $\Delta_L$, and smaller one, $\Delta_S$, upon doping present a higher value than the weak-coupling BCS (1.76$k_{B}T_{c}$) gap value, which reflects a tendency for strong coupling effects. This is inconsistent with the  theoretical constraints of the weakly coupled two-band superconductor model in which one gap must be larger than the BCS gap and one smaller~\cite{TB}.

Although, rather large single or multiple gap values were reported in Fe(Se,Te) from specific heat~\cite{JHu}, penetration depth~\cite{HKi}, and ARPES~\cite{Nak} suggesting strong-coupling multiband superconductivity, the pairing symmetry in Fe(Se,Te) is still under debate. Additionally, two independent reports of penetration depth measurements~\cite{L1} and scanning tunneling microscopy~\cite{L2} in Fe$_{1+y}$(Te$_{1-x}$Se$_{x}$) have claimed the possibility of nodes in the SC gap. Interestingly, near optimal doping FeSe$_{0.45}$Te$_{0.55}$, specific heat measurements demonstrate isotropic gap behavior under zero magnetic field but anisotropic/nodal gaps under magnetic field~\cite{JHu,Nat}. Our data show that Fe(Se,S) system belongs to the class of multiband superconductors, in the strong-coupling regime. Given the substantial divergency of the existing data on the gap values and the gap symmetry for FeSe-based superconductors, a combination of several independent techniques rather than single technique is highly desirable. In the current paper we presented self-consistent data obtained from both lower critical field $H_{c1}$, and specific heat measurements. We believe that other techniques such as $\mu SR$, ARPES, and NMR are highly desirable to further confirm the multiband structure in Fe(Se,S).

\section{Conclusions}

In summary, using a AlCl$_{3}$/KCl flux technique we have grown high quality single-crystals of FeSe$_{1-x}$S$_{x}$ system ($x$=0, 0.04, 0.09, and 0.11) and studied their transport, magnetic and low temperature specific heat properties. We show that the introduction of S to FeSe enhances the upper critical field $H_{c2}$, critical current density $J_{c}$, and the $T_{c}$. The magnetic phase diagram has been studied in the case of magnetic field applied along the  $c$ axis and $ab$ plane for $x$ = 0.04 and the resulting anisotropy was found to be around $\Gamma= H_{c2}^{(ab)}/H_{c2}^{(c)}\sim$ 4. The temperature dependence of the penetration depth and C$_{el}$ can be described neither within single band weak coupling BCS nor using the $d$-wave approach.
Our results, (i) the $T$-dependencies of both penetration depth and specific heat, (ii) the kinky in $\lambda _{ab}(T)$, (iii) the large specific heat SC gap values revealed from both probes, (iv) the linear field dependence of $\gamma$, (vi) the large jump at $T_{c}$, and (vii) the $s$-wave scaling of the low-$T$ specific heat data in the mixed state, all indicate the presence of  strong-coupling multiband and nodeless superconductivity in FeSe$_{1-x}$S$_{x}$. The field-induced change in the low-$T$ specific heat shows a linear magnetic field dependence which is consistent with the $s$-wave symmetry of the order parameter.

\begin{acknowledgments}
We thank S. Hirai, C. Krellner, S. A. Kuzmichev, H. Rosner, S. L. Drechsler, L. Boeri, R. Klingeler, and A. Silhanek for discussions. The work in Russia was supported by Russian Scientific Foundation (Grants Nos.14-13-00738 and 13-02-01180) and the Ministry of Education and Science of the Russian Federation in the framework of Increase Competitiveness Program of NUST (MISiS) (K2-2014-036). G. K. acknowledges the support by a grant from the U.S. Civilian Research and Development Foundation (CRDF Global) OISE-14-60109-0. The work in ECNU was supported by the Natural Science Foundation of China (No. 61125403) and the Ministry of Science and Technology of China (973 projects: Grants Nos. 2013CB922301 and 2014CB921104).
\end{acknowledgments}

\end{document}